\documentclass[aps,showpacs,preprintnumbers,amsmath,amssymb,endfloats]{revtex4}

\usepackage{graphicx}
\usepackage{dcolumn}
\usepackage{bm}

\begin{document}


\title{Superconducting Properties of MgCNi$_3$ Films}

\author{D.P. Young, M. Moldovan, D.D. Craig, and P.W. Adams}
\affiliation{Department of Physics and Astronomy\\Louisiana State University\\Baton Rouge, Louisiana,
70803}%

\author{Julia Y. Chan}
\affiliation{Department of Chemistry\\Louisiana State University\\Baton Rouge, LA, 70803}%

\date{\today}

\begin{abstract}
We report the magnetotransport properties of thin polycrystalline films of the recently discovered
non-oxide perovskite superconductor MgCNi$_3$.  CNi$_3$ precursor films were deposited onto sapphire substrates and
subsequently exposed to Mg vapor at 700 $^o$C.  We report transition temperatures (T$_c$) and critical field values
(H$_{c2}$) of MgCNi$_3$ films ranging in thickness from 7.5 nm to 100 nm.  Films thicker than $\approx40$ nm have a
T$_c\sim 8$ K, and an upper critical field $H_{c2}(T=0)=14$ T, which are both comparable to that of polycrystalline
powders.  Hall measurements in the normal state give a carrier density, $n=-4.2$ x $10^{22}$ cm$^{-3}$, that is
approximately 4 times that reported for bulk samples.   
\end{abstract}

\maketitle

	Over the last decade a broad and significant research effort has emerged aimed at identifying and
characterizing relatively low-T$_c$ superconductors that are exotic in their normal state properties and/or
order parameter symmetries.  Examples include the ternary borocarbides LnNi$_2$B$_2$C, where Ln is a lanthanide
element \cite{Borocarbides}, the non-copper layered perovskite Sr$_2$RuO$_4$ \cite{SrRuO}, and the superconducting
intinerate ferromagnets UGe$_2$ \cite{UGe2} and ZrZn$_2$ \cite{ZrZn2}. The recently discovered intermetallic
MgCNi$_3$ \cite{MgCNi3} falls into this class in that its major constituent, Ni, is ferromagnetic, generating
speculation that the system may be near a ferromagnetic ground state \cite{FerroGS}.  In addition, MgCNi$_3$ is the
only known non-oxide perovskite that superconducts, and is thus a compelling analog to the high-$T_c$ perovskites. 
Notwithstanding the widespread interest in MgCNi$_3$, its status as a non-BCS superconductor remains controversial
\cite{Lin,Voelker}.  Electron tunneling studies of the density of states in polycrystalline powders have yielded
conflicting results as to whether or not MgCNi$_3$ exhibits a BCS density of states spectrum \cite{Mao,Kinoda}. 
Tunneling into sintered powders is technically difficult, and indeed, a detailed quantitative characterization of
MgCNi$_3$ has, in part, been hampered by the fact that only polycrystalline powder samples have been available. 
Obviously, single crystal samples and/or polycrystalline films would be a welcome development both in terms of
fundamental research and possible applications. In the present Communication we present
magnetotransport studies of thin MgCNi$_3$ films.  We show that the transition temperature of the films is only
weakly dependent upon film thickness for thicknesses greater than 10 nm, and that both the transition temperature and
critical field values of films with thicknesses greater than 40 nm are comparable to that of powder samples
synthesized via standard solid state reaction processes. 

	The MgCNi$_3$ films were grown by first depositing thin films of the metastable intermetallic CNi$_3$ onto
sapphire substrates by electron-beam evaporation of CNi$_3$ targets.  Typical deposition rates were $\sim0.1$ nm/s in
a $2$ $\mu$Torr vacuum.  All of the evaporations were made at room temperature, and the resulting films were
handled in air.  The targets consisted of arcmelted buttons of high purity graphite (Johnson Matthey,
$99.9999\%$) and nickel (Johnson Matthey,
$99.999\%$).  The buttons were made with a starting stoichiometry of CNi$_{3.25}$ to compensate for some loss of
nickel during the melting process.  The CNi$_3$ structure of the pristine films was verified by x-ray diffraction. 
Scanning electron microscopy showed the CNi$_3$ films to be very smooth with no discernible morphological features
in 10 $\mu$m x 10 $\mu$m micrographs.  The films were also quite adherent, and could not be pulled off with Scotch
tape.  
	
	MgCNi$_3$ was synthesized by first sealing pristine CNi$_3$ films in a quartz tube under vacuum with approximately
0.1 g of magnesium metal (Alfa Aesar, $99.98\%$).  The tube was then placed in a furnace at 700 $^o$C for 20 minutes,
after which the entire tube was quenched-cooled to room temperature.  X-ray powder diffraction analysis of the
magnesiated films verified that MgCNi$_3$ was formed.  Intensity data were collected using a Bruker Advance D8 powder
diffractometer at ambient temperature in the $2\theta$ range between 20 $\deg$ and 60 $\deg$ with a step width of
0.02 $\deg$ and a 6 s count time.  The inset of Fig.\ 1 shows the X-ray diffraction data for a 90 nm film on a
sapphire substrate.  The powder pattern shows that the film has good crystallinity and that it can be indexed
according to the P{\it m}$\bar{3}${\it m} space group, with $a = 0.38070(2)$ nm.  The pattern also indicates that the
films grew preferentially along the ({\it h}00) reflections.  Electrical resistivity measurements were made by the
standard 4-probe AC technique at 27 Hz with an excitation current of 0.01 mA.  Two-mil platinum wires were attached
to the films with silver epoxy, and the measurements were performed in a 9-Tesla Quantum Design PPMS system from 1.8
- 300 K.

	In the main panel of Fig.\ 1 we plot the resistivity of a 7.5 nm and a 60 nm film as function of temperature in
zero magnetic field.  The thickness values refer to that of the CNi$_3$ layers as determined by a quartz crystal
deposition monitor.  Subsequent profilometer measurements of the magnesiated films did not show any significant
increase in film thickness.  We note that the resistance ratio, $\rho_{290K}/\rho_{10K}\approx3$ of the 60 nm film,
is slightly better than that reported for pressed pellet samples \cite{MgCNi3,Li}.  Indeed, the overall shape of the
60 nm curve is very similar to that of MgCNi$_3$ powders, but the normal state resistivity $\rho_{10K}\approx 20$
$\mu\Omega$-cm, is a factor of 2-6 {\it lower} than polycrystalline powder values.  The midpoint of the resistive
transitions in Fig.\ 1 are 8.2 K and 3.9 K for the 60 nm and 7.5 nm films, respectively.  In Fig.\ 2 we plot the
resistive transitions for a variety of film thicknesses, $d$.  Note that the transition temperature $T_c$ is
relatively insensitive to $d$ down to about $d=15$ nm, below which $T_c$ is suppressed and broadened.

	The perpendicular critical field behavior of a 60 nm film is shown in Fig.\ 3.  As is the case with pressed pellets
of MgCNi$_3$ powder, the resistive critical field transition width is only weakly temperature dependent \cite{Li}. 
We defined the critical field, $H_{c2}$, as the mid-point of the transitions in Fig.\ 3, and in Fig.\ 4 we plot
critical values as a function of temperature.  The solid symbols represent a 60 nm film, and the open symbols are for
a polycrystalline sample from Ref.\ 11.  Clearly the 60 nm critical field behavior is almost identical to that of
sintered MgCNi$_3$ powders, which are known to have an anomalously high critical field and excellent flux pinning
properties \cite{Pinning}.  Future studies of the critical current behavior of the films should prove interesting.

	We have also made Hall measurements of the films in the normal state between 10 K and room temperature.  In the
inset of Fig.\ 4 we show the Hall voltage as a function of magnetic field at 10 K and 200 K.  The solid lines are
linear fits to the data.  The slopes of the lines are proportional to the Hall coefficient
$R_H=1/en$, where $n$ is the effective carrier density.  Clearly the MgCNi$_3$ films have electron-like carriers as
is the case in the bulk material.  However, the calculated carrier density at 10 K, $n=-4.2$ x $10^{22}$ cm$^{-3}$,
is about 4 times larger than that reported in Ref.\ 11 for sintered powder samples.  If, indeed, the carrier density
of the films is enhanced over the bulk values, then it is somewhat surprising that the superconducting properties
remain essentially unchanged relative to that of the bulk material.   

	In conclusion, we report the first synthesis of MgCNi$_3$ films and find that their transition temperatures and
critical field behavior are very similar to that of bulk powder samples.  The films are smooth, adherent, and show
no significant air sensitivity.  A thin film geometry lends itself quite well to planar counter-electrode tunnelling
measurements of the electronic density states, thus providing a compelling alternative to scanning
electron microscopy tunneling.  Such a study in MgCNi$_3$ films should prove invaluable in
resolving the nature of the superconducting condensate. The film geometry will also allow access to the spin
paramagnetic limit in parallel magnetic field studies, as well as possible electric field modulation of the carrier
density via gating.

	We gratefully acknowledge discussions with Dana Browne, Phil Sprunger, and Richard Kurtz. This work was supported by
the National Science Foundation under Grants DMR 01-03892.  We also acknowledge support of the
Louisiana Education Quality Support Fund under Grant No. 2001-04-RD-A-11.

\newpage
\begin{figure}
\includegraphics[width=5in]{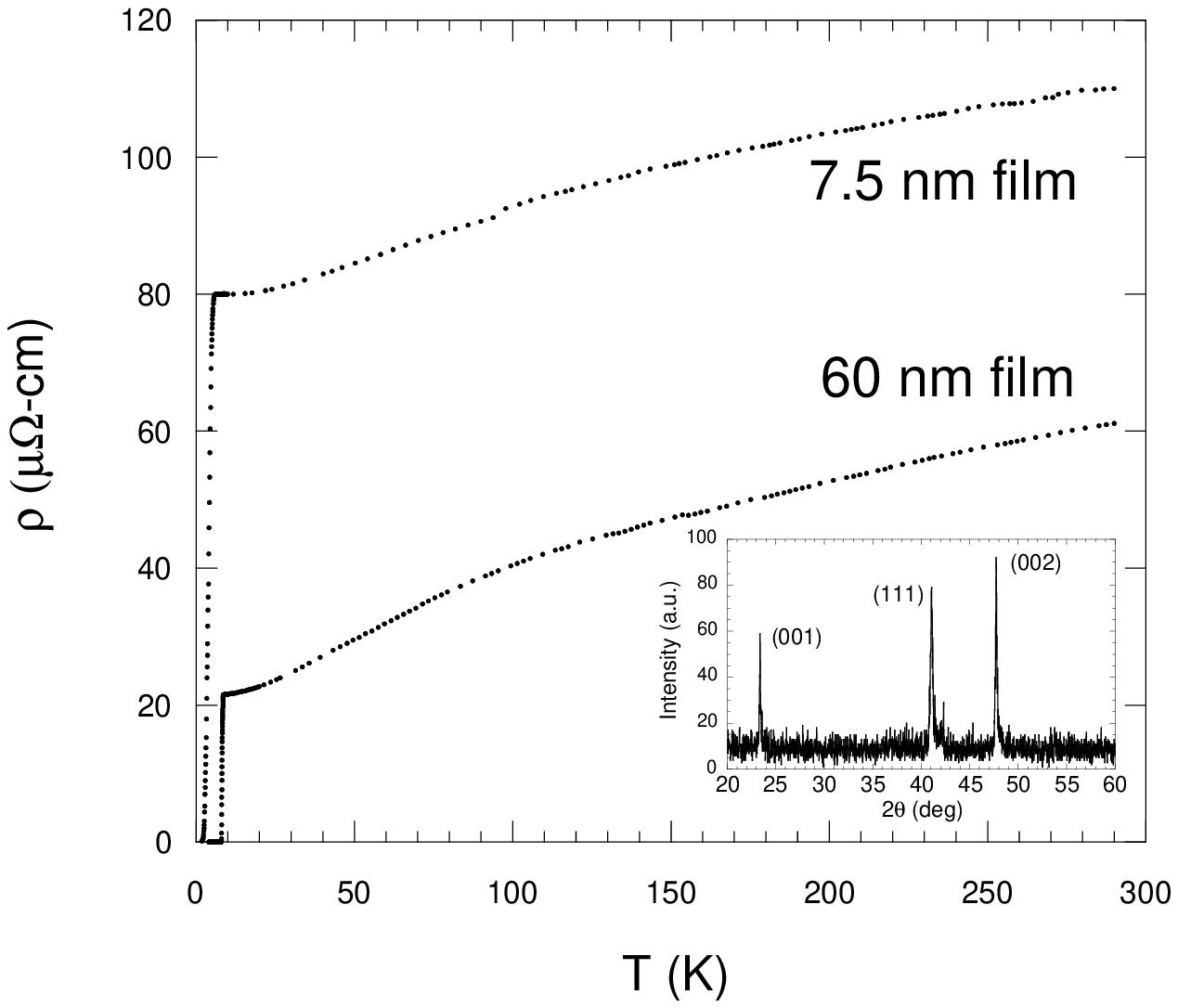}
\caption{\label{fig:epsart} Resistivity of a 60 nm and a 7.5 nm
MgCNi$_3$ film as a function of temperature in zero magnetic field.  The midpoint transition
temperatures of the 60 nm and 7.5 nm films were $T_c=8.2$ K and
$T_c=3.9$ K, respectively.  Inset: X-ray powder diffraction pattern of a 90 nm MgCNi$_3$ film on sapphire.}
\newpage
\end{figure}

\begin{figure}
\includegraphics[width=5in]{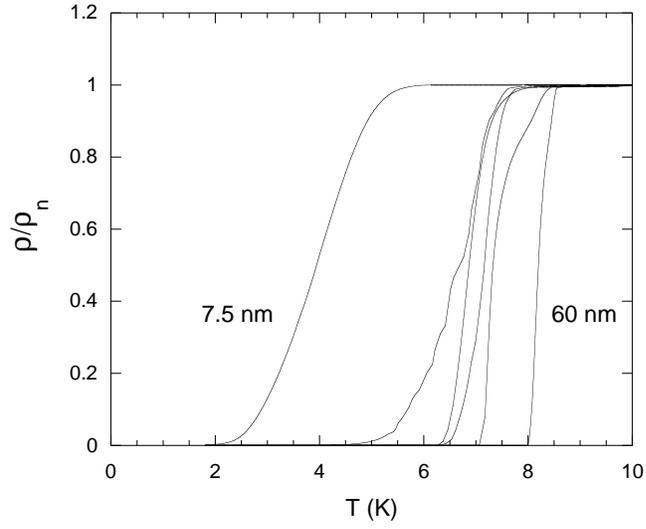}
\caption{\label{fig:epsart}  Resistive transitions for varying film thickness.  The curves correspond from left to
right to MgCNi$_3$ layer thicknesses of 7.5 nm, 15 nm, 30 nm, 45 nm, and 60 nm.}
\newpage
\end{figure}

\begin{figure}
\includegraphics[width=5in]{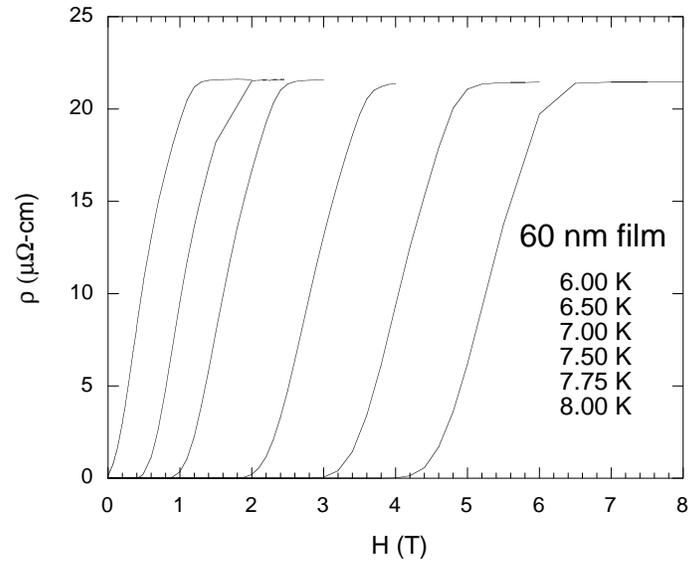}
\caption{\label{fig:epsart} Resistive critical field transitions of a 60 nm film at different temperatures.  The
magnetic field was applied perpendicular to the film surface.}
\newpage
\end{figure}

\begin{figure}
\includegraphics[width=5in]{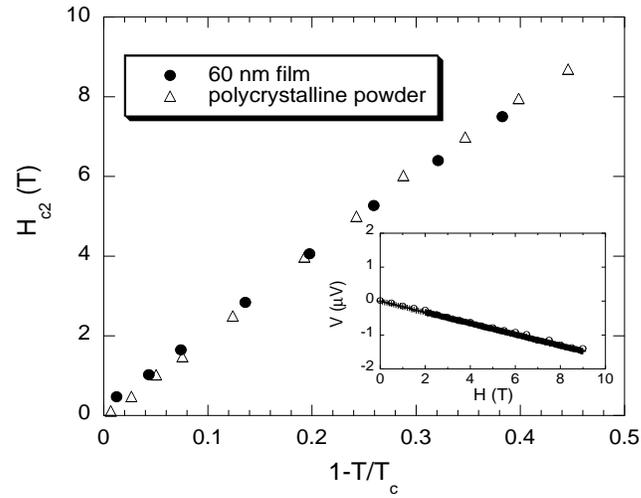}
\caption{\label{fig:epsart} Upper critical field values of the 60 nm film of Fig.\ 3 (solid symbols) and a
polycrystalline sample (open symbols) as a function of reduced temperature.  Inset: Hall voltage 
of a 90 nm MgCNi$_3$ film using a 0.1 mA probe current at 10 K (crosses) and 200 K (circles).  The solid
lines are linear least-squares fits to the data.  The low temperature data corresponds to a carrier density of $n =
-4.2$ x $10^{22}$ cm$^{-3}$.}
\newpage
\end{figure}

\end{document}